\RequirePackage{fix-cm}

\documentclass[smallextended]{svjour3mod}

\smartqed

\usepackage{graphicx}

\usepackage{natbib}
\usepackage{amsmath}
\usepackage{amssymb}
\usepackage[boxed]{algorithm2e}

\newcommand{\R}{\mathbb{R}}
\newcommand{\iid}{\stackrel{\mathrm{i.i.d.}}{\sim}}
\newcommand{\E}{\mathbb{E}}
\newcommand{\Ss}{\mathcal{S}}
\newcommand{\eps}{\varepsilon}
\DeclareMathOperator{\Cov}{Cov}

\journalname{Computational Statistics}

\begin{document}

\title{Sensitivity Analysis for Inference with Partially Identifiable
Covariance Matrices}

\author{Max Grazier G'Sell         \and
        Shai S. Shen-Orr              \and
        Robert Tibshirani
}

\institute{M. Grazier G'Sell \at
              Department of Statistics, Stanford University \\
              \email{maxg@stanford.edu}
           \and
           S. S. Shen-Orr \at
              Department of Immunology, Rappaport Institute for Medical
              Research, \\
              Bruce Rappaport Faculty of Medicine, Technion, Haifa,
              Israel
           \and
           R. Tibshirani \at
              Department of Health Research and Policy, Stanford University\\
              Department of Statistics, Stanford University
}

\maketitle

\begin{abstract}
  In some multivariate problems with missing data, pairs of variables exist
  that are never observed together.  For example, some modern biological tools
  can produce data of this form.  As a result of this structure, the covariance
 matrix is only partially identifiable, and point estimation requires that
 identifying assumptions be made.  These assumptions can introduce an unknown
 and potentially large bias into the inference.  This paper presents a method
 based on semidefinite programming for automatically quantifying this potential bias
 by computing the range of possible equal-likelihood inferred values for convex
 functions of the covariance matrix.  We focus on the bias of missing value
 imputation via conditional expectation and show that our method can give an
 accurate assessment of the true error in cases where estimates based on
 sampling uncertainty alone are overly optimistic.


\end{abstract}

\section{Introduction}
\label{intro}

Methods for estimation in the presence of missing data, particularly the
Expectation-Maximization (EM) Algorithm \citep{littlerubinbook} have seen much development and
use in the last several decades.  One special case are data that arise from mechanisms where some pairs of
variables are never observed together.  This can occur in some modern
biological assays, like multi-dimensional flow and mass cytometry experiments \citep{bendall2012deep}, as well as in more
classical setups like file matching \citep{rubin1986}.
Because these variables are never
observed together, their partial correlation conditional on the other variables
in the data set cannot be estimated.  This leads to a covariance matrix that is only partially
identifiable.

This issue has been addressed in several places in the literature.  In the context of
the EM algorithm, it is suggested that the partial correlations that are
unobservable due to missingness should be
assumed equal to zero \citep{Hartley1971}
\citep{Beale1975}.  In practice, this may be a difficult assumption to
observe.  Furthermore, for modern biological applications where we believe
that the unobservable partial correlations can be influential (and nonzero), it
is important that we be able to assess the possible effects of these
assumptions and to make robust claims in the presence of these unobservable
quantities.

There has been some discussion in the literature of methods for assessing the
effects of these identifying assumptions, particularly in the
file matching literature.  For example, \cite{rubin1986} and
\cite{moriarity2001} both address this problem.  Both of these papers assess
the sensitivity of their methods to the identifying assumptions by
repeating the inference with many possible proposals for the unobservable
partial correlations between variables that are never observed together.
As both acknowledge, it can be difficult to obtain valid and exhaustive
proposals to consider.  Moriarity's paper outlines one approach for obtaining
them; however, it is limited to special cases that are unlikely to occur in the
biological data we are considering. For large, complicated problems, the set of possible
partial correlations can be quite complicated, and an automatic method for
handling the ambiguity over this set would be helpful.

In the next section, we describe the details of our setup and of existing
approaches, and discuss the issues and concerns that arise due to the partial
identifiability of the covariance matrix.  In Section \ref{proposal}, we
propose a method for automatically conducting sensitivity analysis in this setting using
semidefinite programming.  In Section \ref{results}, we
present simulation results and an example on biological data.  Finally, in
Section \ref{discussion}, we discuss the method, its shortcomings, and possible
extensions.

\section{Setup and  Existing Approaches}\label{exist}

In this section, we describe the setup and notation for our problem along with
the existing estimation procedures that exist for data of this form.  We discuss some of the concerns that arise in these procedures due to
partial identifiability, which we address in later sections with our proposed method.  Finally, we present an example and a simulation of the scenario we are discussing.

\subsection{Setup and Notation}\label{exist:setup}

Suppose that we have latent observations $X_i^* \in \R^p$, $i=1,\dots,n$, some of which may be
unobserved.  We will assume for this paper that
\begin{equation}\label{eq:setup1}
  X_i^* \iid N(\mu,\Sigma),\qquad \mu\in\R^p,\qquad \Sigma\in\R^{p\times p}.
\end{equation}

\begin{figure}[ht!]
  \begin{center}
    \includegraphics[width=.75\linewidth]{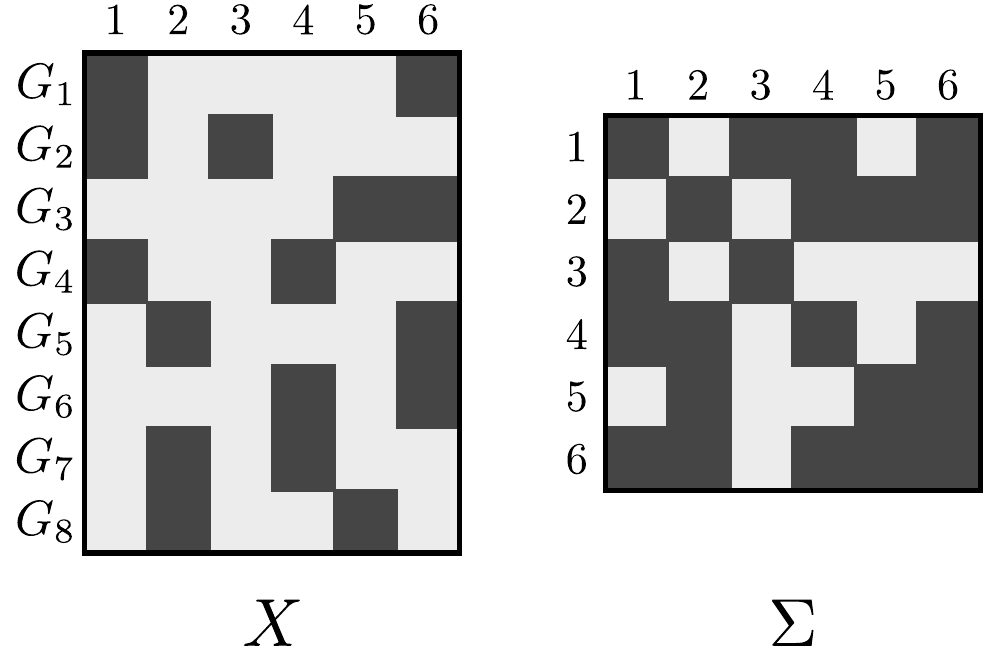}
  \end{center}
  \caption{A representation of the data setup.  The rows of $X$ correspond to
  observations, while the columns of $X$ correspond to variables.  Both indices
  of $\Sigma$ correspond to variables.  Observed entries of $X$ are
  shown in black, as are the corresponding observable entries of $\Sigma$.  The
  light gray regions of $\Sigma$ cannot be determined based on the data, since those
  pairs never appear together in the same measurement block $G_k$.  The indices
1 through 6 are provided to show the correspondence between the data matrix and
the covariance matrix.}
  \label{matrixsetup}
\end{figure}

The observations are broken up into blocks $G_1,\dots,G_K$ that form a
partition of $\{1,\dots,n\}$.  Within block $G_k$, only variables $J_k\subset
\{1,\dots,p\}$ are actually measured.  That is, our observed variable
$X\in\R^{n\times p}$ is given by
\begin{align}\label{eq:setup2}
  X_{ij} &= \begin{cases}
    X_{ij}^* & (i,j) \in \bigcup_k \left(G_k\times J_k\right)\\
    \mathrm{NA} & \text{otherwise}
  \end{cases},
\end{align}
where $\mathrm{NA}$ refers to an observation that is censored or missing.  A
graphical representation of this setup is shown in Figure \ref{matrixsetup}.

The case described in Equations (\ref{eq:setup1}) and (\ref{eq:setup2}) is one
for which the Expectation-Maximization (EM) Algorithm is
commonly used.  The algorithm is detailed in \cite{littlerubinbook}, and described briefly in the next
subsection.

We are interested in the particular case where there exist pairs of variables $j_1,j_2 \in\{1,\dots,p\}$ such that there
is no $J_k$ containing both $j_1$ and $j_2$.  This will result in inestimable
partial correlations between those variables, leading the final covariance matrix
to be under identified.  This is discussed further in
Section \ref{exist:concern}.

It is worth noting that another property of several of the data sources we are
interested in, particularly the
mass cytometry data discussed in Section \ref{results:data}, is that the number of
observations in each block $G_k$ is very large, so that the parameters
corresponding to the individual blocks can be very well determined.  As a result, sampling
noise turns out to be less important than the partial identifiability of the
covariance matrix.

\paragraph{Note on notation for indices.}  In several parts of this paper, we will be
referencing sub-blocks of $\Sigma$.  In those cases, indexing by sets like
$J_k$ will refer to all of the corresponding elements.  For example,
$\Sigma_{J_k,J_k}$ refers to the sub-matrix with row and column indices in $J_k$,
while $\Sigma_{j,J_k}$ refers to the elements of row $j$ with columns in $J_k$.

\subsection{Estimation Approaches}\label{exist:est}

The standard approach for missing data problems like these is to maximize
the observed log-likelihood \cite{littlerubinbook}.  In this case, the observed
log-likelihood is given by
\begin{align*}
  \ell_{obs}(X,\mu,\Sigma) &= \mathrm{const} - \frac{1}{2}\sum_{k=1}^K \sum_{i
  \in G_k} \log\left|\Sigma_{J_k,J_k}\right|\\
&- \frac{1}{2}\sum_{k=1}^K \sum_{i \in G_k}
  \left(X_{i,J_k}-\mu_{J_k}\right)^T\Sigma_{J_k,J_k}^{-1}\left(X_{i,J_k}-\mu_{J_k}\right)
\end{align*}
which is just the sum of the marginal log-likelihoods for each of the
observed multivariate normal vectors.

We note here, and discuss further in Section \ref{exist:concern},
that this observed log-likelihood depends only on those coordinates of $\Sigma$
corresponding to variables that are observed together in some block $G_k$.  In
our notation, it depends only on elements $\Sigma_{ij}$ where $(i,j)\in \bigcup_k \left(J_k\times
J_k\right)$.

A traditional approach to maximizing this observed log-likelihood has been
the EM Algorithm, described in several places, including detailed coverage
in \cite{littlerubinbook}.  For
review, we include a quick summary from their book, recast in our notation.

Let $X^{(t)}, \mu^{(t)}, \Sigma^{(t)}$ be the estimates at stage $t$ (where
$X^{(t)}$ will no longer be missing, since we will impute the values).  We begin
by initializing $\mu^{(1)},\Sigma^{(1)}$.  Several methods for initialization
are discussed in \cite{littlerubinbook}.  Then an iteration of the algorithm
progresses as:

\noindent E-Step:
\begin{align*}
  X_{ij}^{(t)} &= \begin{cases}
    X_{ij} & (i,j)\in\bigcup_k \left(G_k\times J_k\right) \qquad\text{(observed)}\\
    \E(X_{ij}|X,\mu^{(t)},\Sigma^{(t)}) &  (i,j)\notin\bigcup_k \left(G_k\times J_k\right)
    \qquad\text{(unobserved)}\\
  \end{cases}
\end{align*}
\noindent M-Step:
\begin{align*}
  \mu_j^{(t+1)} &= \frac{1}{n}\sum_{i=1}^n X_{ij}^{(t)}\\
  \Sigma_{jk}^{(t+1)} &= \frac{1}{n} \sum_{i=1}^n \left(\left(X_{ij}^{(t)} -
  \mu_j^{(t+1)}\right)\left(X_{ik}^{(t)}-\mu_k^{(t+1)}\right)+c_{jki}^{(t)}\right)\\
  c_{jki}^{(t)} &= \begin{cases}
    0 & \text{$X_{ij}$ or $X_{ik}$ observed}\\
    \Cov\left(X_{ij},X_{ik}|X,\mu^{(t)},\Sigma^{(t)}\right) &
    \text{$X_{ij}$ and $X_{ik}$ both missing}
  \end{cases}
\end{align*}

Here, conditioning on the matrix $X$ (not $X^{(t)}$) is being used to represent
conditioning on the observed, uncensored elements of the actual data $X$.  The
$c_{jki}^{(t)}$ terms in the algorithm adjust for low covariance estimates in cases
where both data coordinates are missing from $X$.  This is because using
the imputed means for both coordinates would lead to inaccurately low
covariance estimates.

Note that once $\mu^{(1)}$ and $\Sigma^{(1)}$ are initialized, execution of this
algorithm will not be affected by the presence of partial identifiability.  The previous
literature addresses this problem of partially identified $\Sigma$, suggesting that
the covariance matrix be initialized with the unobservable partial correlations
set to zero. \citep{Hartley1971} \citep{Beale1975}

For certain structures of missing data, procedures for restricting these
unobservable partial correlations to zero exist in the literature.  This is
true for factorizable designs as in \cite{littlerubinbook}, where the SWEEP
operator is used to make the operation convenient.  To the best of our
knowledge, no procedure for restricting the unobservable partial correlations
to zero in the EM algorithm has been set out in the literature for general
designs. However, regardless of the assumptions used to make the estimation
identifiable, a set of equal-likelihood estimates will exist and it is important to
understand how selection among them could impact inference.

\subsection{Concerns about Partial Identifiability}\label{exist:concern}

As noted in Section \ref{exist:setup}, the observed log-likelihood depends on $\Sigma$
only through those elements that appear together in measurement sets
$J_1,\dots,J_K$.  Suppose that $\hat{\Sigma}$ is an estimate of the covariance
matrix, in this case the result from the EM algorithm.  Consider the set
\begin{equation}
  \Ss = \left\{\Sigma \succeq 0: \Sigma_{ij} = \hat{\Sigma}_{ij} \text{ for }
  (i,j)\in\bigcup_k \left(J_k\times J_k\right)\right\},
\end{equation}
where $\Sigma \succeq 0$ indicates $\Sigma$ in the positive semidefinite
cone.  Because all the matrices in this set are identical on the coordinates
that appear in the observed log-likelihood, they all correspond to the same
value of the observed log-likelihood as the original estimate $\hat{\Sigma}$.  Based
on our data alone, there is no reason to necessarily choose one matrix in $\Ss$ over
another.

Any point estimation procedure, either implicitly or explicitly, includes an
assumption on these missing elements in order to obtain a point estimate, which
will be an element of $\Ss$ by construction.  The
recommended practice of initializing $\Sigma^{(1)}$ in the EM algorithm to have
zero for the unobservable partial correlations is one such approach.  Other
initializations will carry with them other implicit assumptions.
These assumptions have the potential to introduce unknown biases into our
estimates of $\Sigma$ and into later inference that depends on these estimates.

The usual assumption for the EM algorithm is quite reasonable, as it attempts
to place the estimate near the middle of the equivalence set $\Ss$.  Nevertheless, we
might wonder what effect the width of that set could have on our error.  In
particular, we might wonder how these unknown biases affect
inference based on $\hat{\Sigma}$.  The most common such inference is
imputation of the missing elements of $X$ by conditional expectation.  These
values are returned as part of the usual EM algorithm.

The assumptions required to identify $\hat{\Sigma}$ among the elements of $\Ss$
could lead to an unknown bias in the imputed conditional expectations.  Since
this unknown bias would not be captured by sampling uncertainty, usual error
estimation practices like bootstrapping could be misleadingly optimistic about
the accuracy of our imputations.

\subsection{Simple $3\times 3$ Example}\label{exist:example}

To illustrate the type of partial identification we are discussing, consider
the following $3\times 3$ example.  This example is much simpler than the
experimental settings we intend to address, but it gives a clearer
understanding of the set $\Ss$.

\begin{figure}[ht!]
  \begin{center}
    \includegraphics[width=0.5\linewidth]{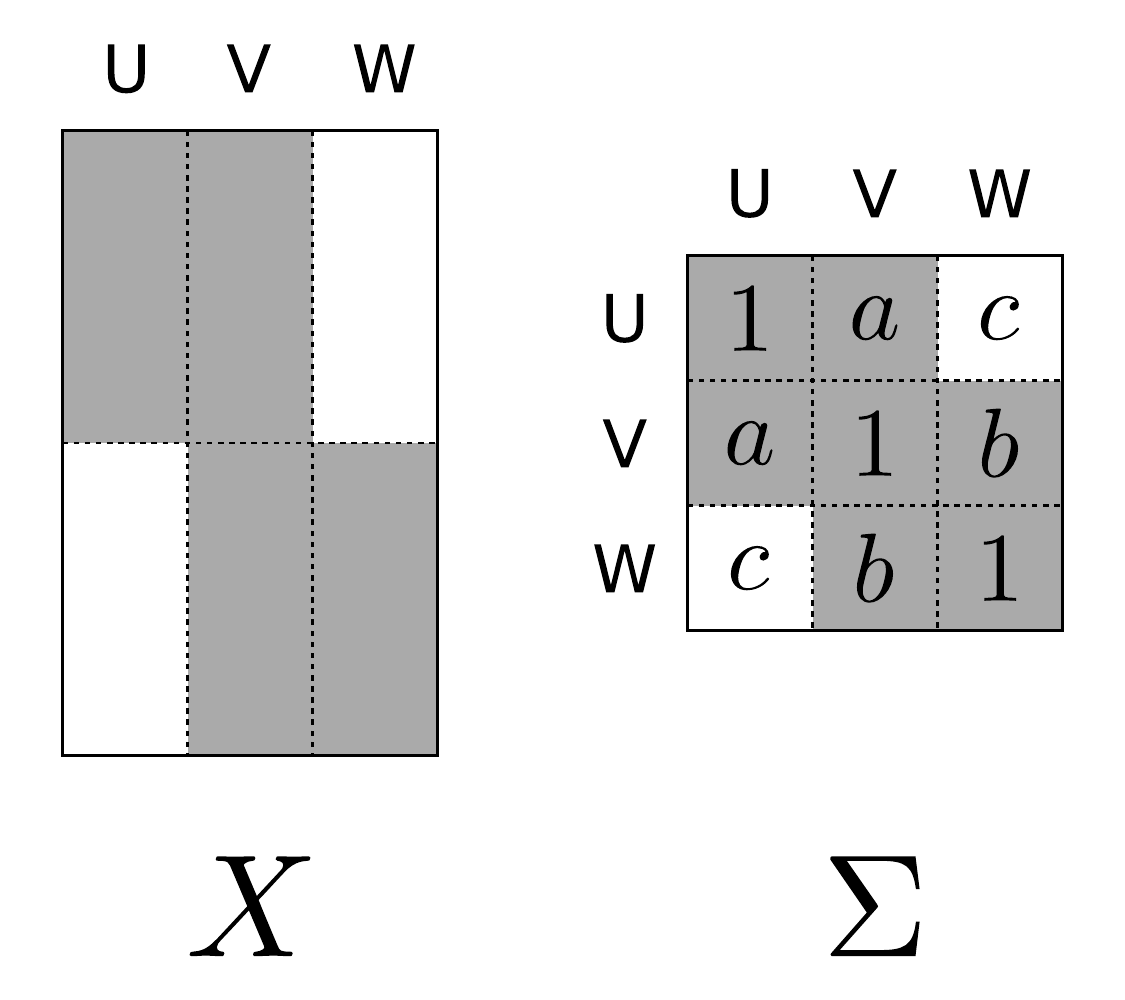}\\
  \end{center}
  \caption{Simple data structure illustrating partial identification.
  Variables $U$ and $W$ are never observed together, leading the unshaded
  element of $\Sigma$ unidentifiable.}
  \label{threeexample}
\end{figure}

Suppose that $p=3$, and that $X$ is made up of only two blocks of measurements.
Let $U,V$ and $W$ be the three variables in $X$.  We assume the structure shown
in Figure \ref{threeexample}.

In this structure, the pairs $U,V$ and $V,W$ have been observed together, but
$U$ and $W$ never appear together.  Thus the correlation between $U$ and $W$
cannot be identified, it can only be bounded by the observable correlations.
The only constraint on $c$ is that $\Sigma$ is positive semidefinite.  In
this particularly nice example, we can work this constraint out algebraically
to see that
\begin{equation*}
  c \in \left[ab - \sqrt{\left(1-a^2\right)\left(1-b^2\right)},ab + \sqrt{\left(1-a^2\right)\left(1-b^2\right)}\right].
\end{equation*}
The usual recommendation that the EM algorithm estimate be constrained to zero
for the unobservable correlation is equivalent in this case to constraining $c=ab$.

As a result of the lack of identifiability of $c$, whatever estimates $\hat{a},\hat{b}$ are obtained, the set
\begin{equation*}
  \Ss = \left\{\Sigma: a=\hat{a}, b=\hat{b}, c \in \left[ab - \sqrt{\left(1-a^2\right)\left(1-b^2\right)},ab + \sqrt{\left(1-a^2\right)\left(1-b^2\right)}\right]\right\}
\end{equation*}
has equal observed likelihood, so all the elements are indistinguishable based
on the data alone.  This is the sort of partial identifiability with which we
are concerned.  Any point estimate $\hat{c}$ (for example $\hat{c}=\hat{a}\hat{b}$) corresponding to some
$\hat{\Sigma} \in \Ss$ could introduce an unknown
bias, both into $\hat{c}$ and into any further inference depending on $\hat{c}$.

\subsection{Simulation}\label{exist:simulation}

For a more realistic example, consider the following simulation.  We construct
a simulated data set $X$ drawn from a multivariate normal.  For this setup, we
choose $n=5000$ samples, $p=18$ variables, $K=5$ groups, $|J_k|=12$ variables measured within each group.  Within each group, the 12 observed variables are chosen randomly.  For our realization, this leads to nine pairs of variables
that never appear together (out of 153 total pairs).  Let $\mu=0$ for
convenience.  To make the effect dramatic, let $\Sigma$ be 1 on the diagonal,
$0.3$ for each of the observable entries, and $0.56$  for each of the
unobserved entries (chosen to keep $\Sigma$ positive semidefinite).  Since these unobservable
entries in $\Sigma$ are never measured, we expect any inference fail to take
them into account.  We initialize the EM algorithm with the unobserved entries
set to 0.3, in agreement with all the observed entries.

\begin{figure}[ht!]
  \begin{center}
    \includegraphics[width=.75\linewidth]{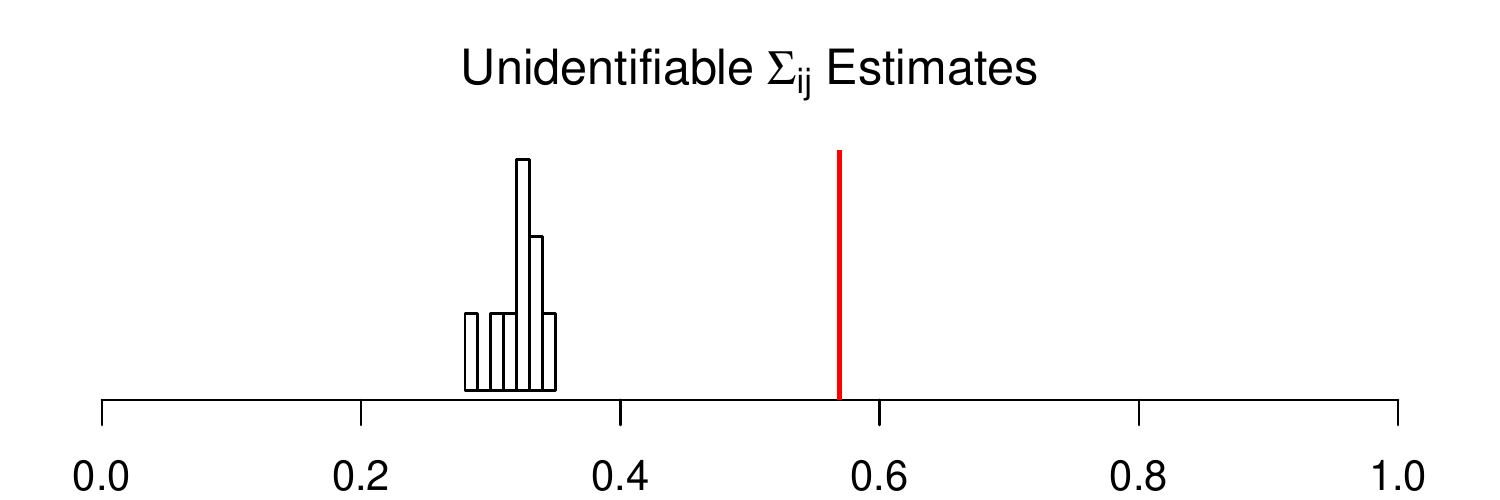}\\
  \end{center}
  \caption{The histogram shows the estimates for the unidentified elements
  of $\Sigma$, while the vertical red line shows the true value in our
  simulation.}
  \label{sigmaerr}
\end{figure}

We run the EM algorithm to estimate $\hat{\Sigma}$.  The plot in Figure
\ref{sigmaerr} shows the
resulting estimates for the unobservable elements of $\Sigma$, along with the
true value.  We see that the estimates miss the true value, as we would expect.

\begin{figure}[ht!]
  \begin{center}
    \includegraphics[width=\linewidth]{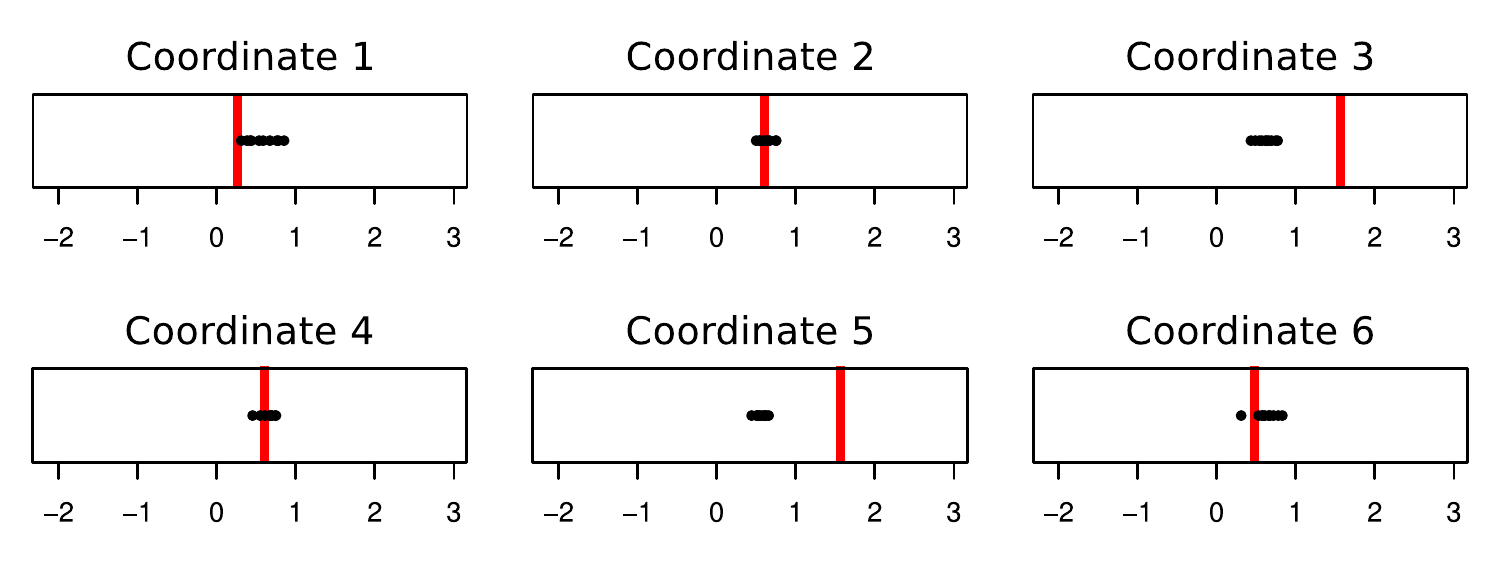}
  \end{center}
  \caption{Each subplot shows the ten imputed values for a missing element of
  $X$, based on ten different realizations of the simulation.  The vertical red
  lines show the true values for those coordinates.  We see that the sampling
  uncertainty in the imputations is very small, but that it does not capture
  the missing bias due to the estimation assumptions.}
  \label{imputeerr}
\end{figure}

As a demonstration, we consider ten realizations of the previous simulation.
For each, we look at the imputed and actual values for the missing coordinates
in the first row of $X$.  These are shown in Figure \ref{imputeerr}.  We see
that the sampling uncertainty does not capture the actual error from the true
value, which is quite large in some cases.  This suggests that, in this
setting, the unknown error due to the ambiguity in $\Ss$ is important to
understand.

In the next section, we propose an approach to determining the
sensitivity to the set $\Ss$ of inference based on $\Sigma$, so that this error
can be understood and bounded.

\section{Sensitivity Analysis by Semidefinite Programming}\label{proposal}

From the previous section, we recognize that because some pairs of variables
are never observed together, any estimate $\hat{\Sigma}$ has a set of
equivalently likely estimates
\begin{equation*}
  \Ss = \{\Sigma \succeq 0: \Sigma_{ij} = \hat{\Sigma}_{ij} \text{ for }
  (i,j)\in\bigcup_k \left(J_k\times J_k\right)\}
\end{equation*}
which are indistinguishable based on the data.  Any point estimate of $\Sigma$
must choose a particular element of $\Ss$, which in turn can lead to an unknown
bias in inference based on $\Sigma$.

Accepting the ambiguity of $\Ss$ as an inherent shortcoming of the lack of coverage of our
measurements, we would still like to be able to capture the range of possible
values in our inference.  We saw in Figure \ref{imputeerr} that the usual
methods for estimating sampling error, like the bootstrap, will not address this
bias.  In this section we propose and describe a method for sensitivity
analysis in this setting, giving bounds on the range of imputed values with
respect to all of $\Ss$.

\subsection{General Approach}\label{proposal:general}

Our goal here is to bound the values obtained by inference on $\Sigma$ over the
entire set $\Ss$.  It is convenient to notice that $\Ss$ is the intersection of
the convex semidefinite cone $\{\Sigma\in\R^{p\times p}: \Sigma \succeq 0\}$ with the affine set
$\{\Sigma\in\R^{p\times p}: \Sigma_{ij} = \hat{\Sigma}_{ij} \text{ for }
(i,j)\in\bigcup_k \left(J_k\times J_k\right)\}$.  As a result, $\Ss$ itself is a convex set.

We are interested in understanding the effect of the under-identified nature of
$\Sigma$ on inferred quantities based on $\Sigma$.  The example given above, and which we will
continue to use, is that of imputation by conditional expectation.  For a
general inferred quantity $f(\Sigma)$ (potentially also a function of $X$ and $\mu$),
we can look at extrema of the function $f$ over $\Ss$.  For the case of
convex or affine $f$, this can be tractable because of the convexity of $\Ss$,
and we will be able to bring to bear machinery from the study of convex
optimization and particularly semidefinite programming.
Finding such extrema will give bounds on the range of possible values of $f$
over the entire set $\Ss$ of likelihood maximizing matrices.

In the next subsection, we expand on this idea for the case where $f$ is
the conditional expectation.  We will return to general applications briefly in
the discussion in Section \ref{discussion}.

\subsection{Algorithm: Conditional Expectation}\label{proposal:algorithm}

In this section, we are interested in applying the idea from Section
\ref{proposal:general} to conditional expectation, particularly the imputations
of the missing values in $X$.  For this section, we fix $i$ and $j$ so $X_{ij}$
is the missing element we would like to impute, and we fix $k$ so that $i \in
G_k$ and $j \notin J_k$.  The quantities of interest then take the form
\begin{equation}
  \label{condexp1}
  f(\Sigma) = \E(X_{ij}|X_{iJ_k}) = \mu_j +
  \Sigma_{jJ_k}\Sigma_{J_kJ_k}^{-1}\left(X_{iJ_k}-\mu_{J_k}\right),
\end{equation}
corresponding to the conditional expectation of the $i^{th}$ row and $j^{th}$ column of
$X$.  This section focuses on understanding the range of possible values for (\ref{condexp1}) for $\Sigma\in\Ss$.

The quantity in (\ref{condexp1}) appears to be non-convex in $\Sigma$.
However, we are interested in optimizing it only over $\Sigma \in \Ss$.  Within
$\Ss$, the block $\Sigma_{J_k,J_k}$ is held constant, since it corresponds to
coordinates that are observed together in block $G_k$.  This means that as long as $\Sigma
\in \Ss$, $\Sigma_{J_k,J_k}$ is fixed and can be set to
$\hat{\Sigma}_{J_k,J_k}$, the corresponding block of our initial estimate
(e.g., from the EM algorithm).  This results in the simplified version of the
objective from Equation (\ref{condexp1}),
\begin{equation}
  \label{condexp2}
  f(\Sigma) = \E(X_{ij}|X_{iJ_k}) = \mu_j +
  \Sigma_{jJ_k}\hat{\Sigma}_{J_kJ_k}^{-1}\left(X_{iJ_k}-\mu_{J_k}\right).
\end{equation}
This is not only convex in $\Sigma$ over $\Ss$, but is affine.  As a result, (\ref{condexp2}) has both unique
maxima and unique minima over $\Ss$.

Finding the range of possible imputed values is then equivalent to solving the
semidefinite program (SDP)

\begin{align*}
  \underset{\Sigma}{\mathrm{[min/max]imize}}&\;\;
  \Sigma_{j,J_k}\hat{\Sigma}_{J_k,J_k}^{-1}(X_{J_k}-\hat{\mu}_{J_k})\\
  \text{subject to } & \;\; \Sigma \succeq 0\\
  & \;\; \Sigma_{ab} = \hat{\Sigma}_{ab} \text{ for $(a,b)\in\bigcup_{\ell}
  \left(J_{\ell}\times J_{\ell}\right)$}
\end{align*}

Here we have dropped the $\mu_j$ term for conciseness, since it is just a constant and can be
reintroduced later.

The algorithm for finding the minimum and the maximum of this affine function
will be the same, so we will just focus on the minimum here.  To solve the SDP,
we will replace the semidefinite cone constraint ($\Sigma \succeq 0$) with a
log-barrier \citep{boyd2004convex}.  As $t$ grows large, the following
minimization problem is equivalent to the original one.
\begin{align*}
  \underset{\Sigma}{\mathrm{minimize}} &\;\;
  -\frac{1}{t}\log|\Sigma| + \Sigma_{j,J_k}\hat{\Sigma}_{J_k,J_k}^{-1}(x_{J_k}-\hat{\mu}_{J_k})\\
  \text{subject to } &\;\; \Sigma_{ab} = \hat{\Sigma}_{ab} \text{ for
  $(a,b)\in\bigcup_{\ell} \left(J_{\ell}\times J_{\ell}\right)$}.
\end{align*}

To solve the original SDP, we solve this minimization problem over a sequence
of increasing $t$ with warm starts (see Appendix A for details). The preceding optimization problem becomes the inner loop of the
algorithm.  Each minimization is computed by generalized gradient descent.
Furthermore, we apply recent approaches for accelerating gradient descent, for example
in \cite{banerjee2006convex} and \cite{beck2010gradient}, to speed up convergence.  The timings below show
that this acceleration provided much faster convergence times.  The details of the algorithm can be found in
the appendix to this paper.

\begin{figure}[ht!]
  \begin{center}
    \includegraphics[width=\linewidth]{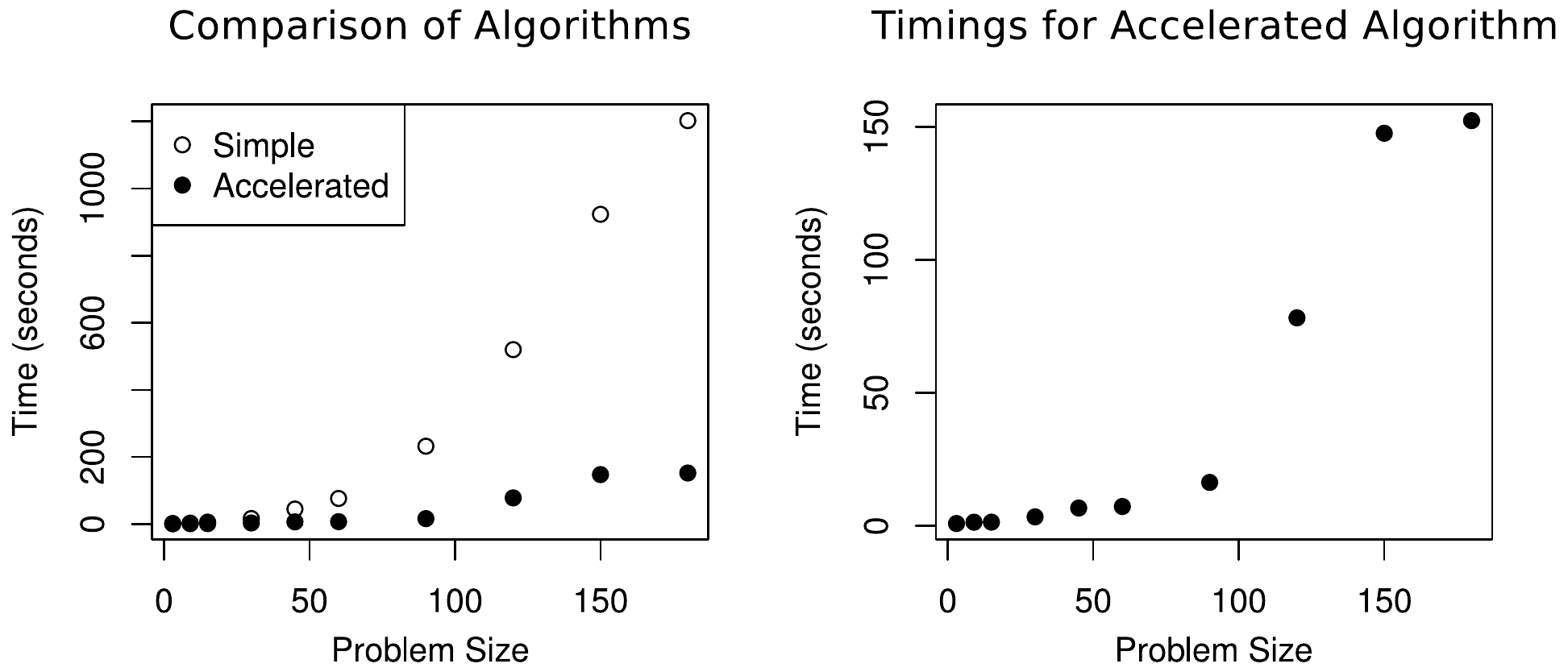}
  \end{center}
  \caption{Time in seconds to compute upper and lower bounds for one imputed
  entry.  This involves executing the SDP algorithm twice.  Each entry is an
  average over ten repetitions.}
  \label{timingplot}
\end{figure}

\paragraph{Timings.}  Timings on a variety of problem sizes are shown in Figure \ref{timingplot}.  A comparison is made between generalized gradient descent with and without acceleration.  These simulations are all run on a 3.3Ghz Intel Xeon X5680 processor.

The limiting computation at each step of the algorithm is a singular value decomposition to compute
the inverse and check the step size conditions, at a cost of
$O(p^3)$.  We see that the computation is possible through $p$ in the low hundreds.  The accelerated form of the algorithm is an order of magnitude
faster on large problems, since it requires fewer iterations to converge to the solution and thus fewer costly matrix decompositions.

It is
worth noting that computing these bounds for all the missing coordinates of $X$ could become quite expensive.
 Conveniently, the algorithm is quite parallelizable, since
the entries of each row of $X$ can be bounded with no information sharing
from other rows.  This makes the algorithm well-suited to being distributed
over large computing clusters.  The authors are interested in pursuing
this possibility to make even large problems accessible for this method.

\section{Results}\label{results}
In this section, we give results from the proposed algorithm for sensitivity
analysis of the imputed conditional expectations.  The first example is the same
simulation setup as outlined in Section \ref{exist:concern} and shown in
Figures \ref{sigmaerr} and \ref{imputeerr}. The second is mass cytometry data, which was the setting that inspired our interest in this
problem.

\subsection{Simulation}\label{results:sim}

The simulation setting is the same as outlined in Section \ref{exist:concern}.
We run the same simulation as in Figure \ref{imputeerr}, looking at the true
and imputed conditional means.  As in the first simulation, we do this for 10
realizations of the data, looking at the six imputed values in the first row of
$X$.

This time, we use the SDP algorithm above to obtain a range of possible imputed
values over all of $\Ss$.  This gives bounds on the bias due to assumptions on
the missing partial correlations.

\begin{figure}[ht!]
  \begin{center}
    \includegraphics[width=\linewidth]{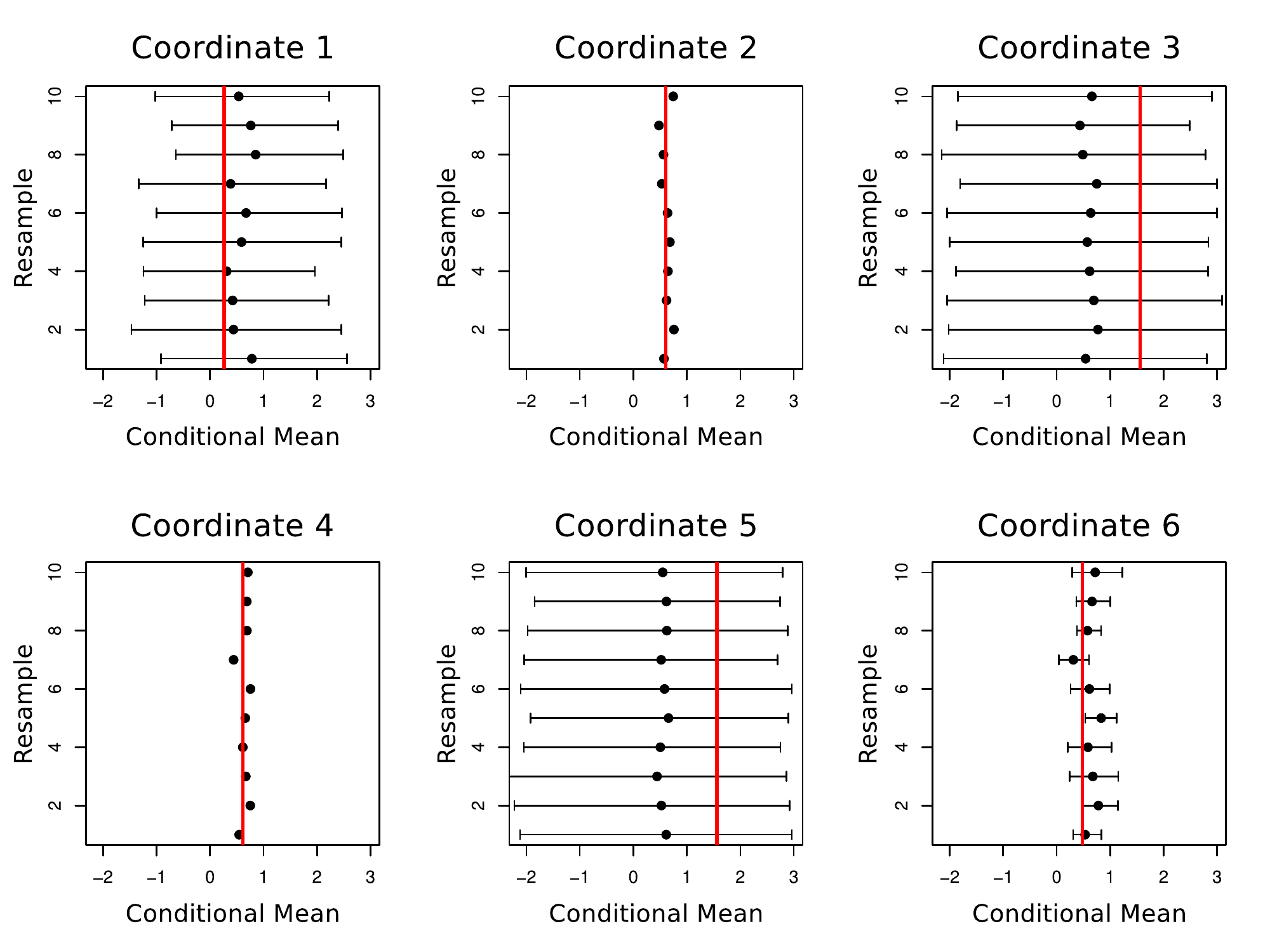}
  \end{center}
  \caption{Imputed values over 10 realizations are shown as black dots.  The
  computed range of possible imputed values due missing partial correlations is shown by black
  intervals.  The vertical red lines are the true conditional means.}
  \label{simintervals}
\end{figure}

We see in Figure \ref{simintervals} that the intervals cover the true values,
capturing the unobservable bias.  The method gives an indication of the
reliability of particular imputations, taking into account how well the
relevant correlations are constrained by the observable ones.  Coordinates 2
and 4 depend only on observable correlations, so their intervals have zero
length since $\Ss$ is degenerate.  Coordinates 3 and 5 have very wide intervals
and are quite conservative, but this is merely an realistic assessment of the
data structure's inability to constrain the correlations of interest.  These
are examples of the wide intervals our method can produce.  While it would be
desirable to have shorter intervals, it is not possible.  The wide intervals
produced by our approach correspond to estimates that are all indistinguishable
based on the data; there is no information in the data to distinguish between
points in the intervals.

In cases like coordinate 1 and particularly coordinate 6, the method gives
quite narrow ranges, indicating that the relevant correlations have been
well-constrained by the estimable correlations, giving reliable imputations.
This is valuable information to have, especially in settings where we expect
strong correlations to occur, like biology.  Our next section gives an example
using mass cytometry data, where data with this missing structure arise
inherently.

\subsection{Real Mass Cytometry Data}\label{results:data}

Our interest in this
area was originally inspired by mass cytometry data from Time-of-Flight (cyTOF)
experiments.  The structure of the data in those experiments lends itself to
the designs we have described in Section \ref{exist:setup}.  The technology is
described in detail in \cite{bendall2012deep}.  We will give a brief overview
below.  We note that though we discuss the ideas presented here in the context
of newer mass cytometry technology, they would in principle be applicable to
the more established fluorescence based flow cytometry technology.

Mass cytometry experiments run on a cyTOF machine measure protein expression of
many proteins on a single cell basis.  This is done by attaching tagged
antibodies to the cells, which bind to specific proteins.  Each protein
specific antibody is tagged with heavy metal atoms of a specified type.  The
cells are then atomized one at a time and sent through a mass spectrometer.
The mass spectrometer measures all of the heavy metal tags present, each
reflecting the abundance of expression of a specific protein in each cell.

The data structure that we have been discussing arises because, while there are
several hundreds of proteins of interest, there are only a limited number of
different heavy metals that can be bound to these antibodies.  As of this
article, the state of the art is 45 unique metal tags \citep{bendall2012deep} and 17
unique fluorescent tags in flow cytometry \citep{flownumber}.

To study more proteins, one could
construct several
different sets of antibody tags, each using the same metals to code for different
proteins.  The cell population can then be divided into
batches, and each can be run against a different set of probes.

\begin{figure}[ht!] \begin{center}
  \includegraphics[width=\linewidth]{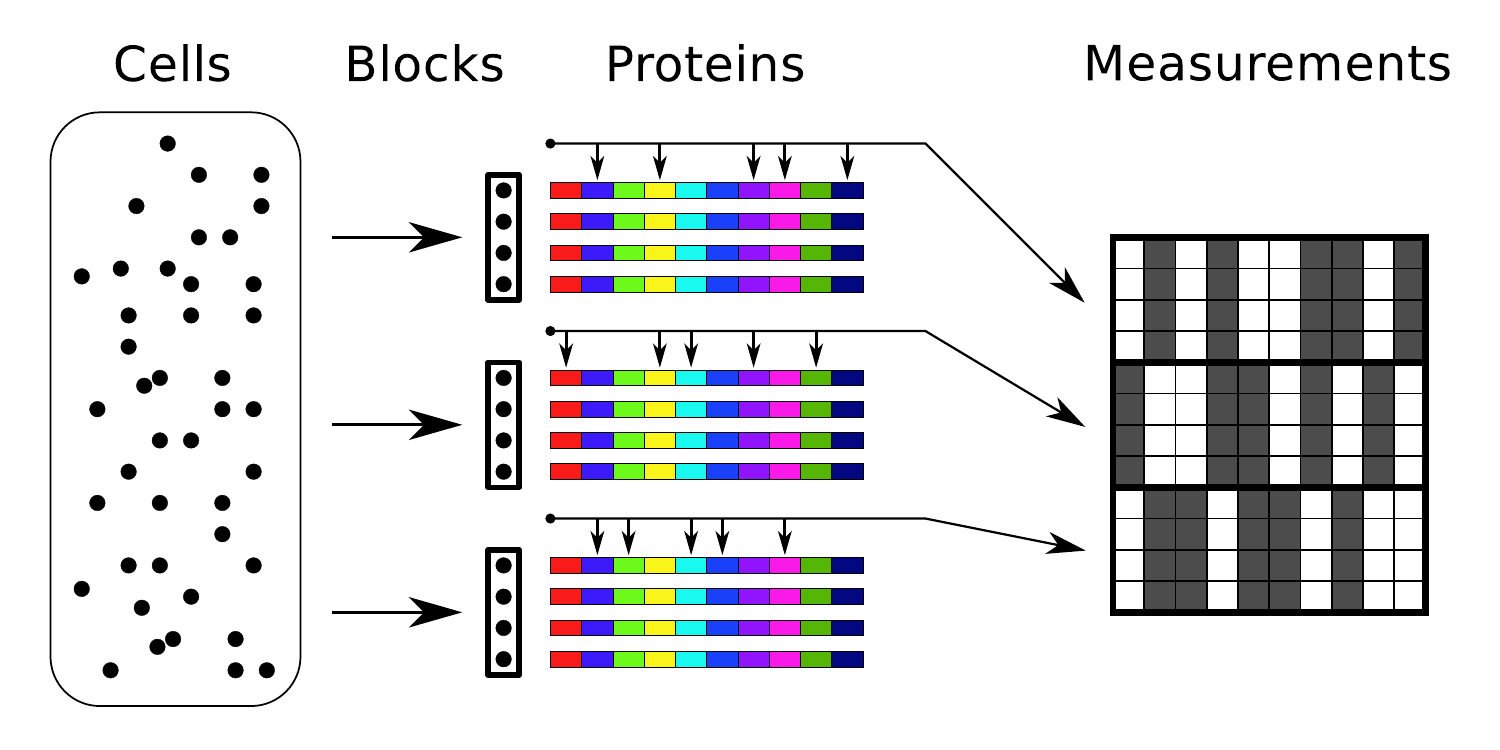} \end{center}
  \caption{Representation of the mass cytometry data acquisition.  The incoming
  cells are grouped into blocks.  Each block is measured with a unique set of
  probes, which measure a subset of the available proteins.  The protein colors
  represent protein identity; although some amount of each protein may be
  present in each cell, the probes can only measure a subset of the protein
  types.  This measurement scheme leads to the data matrix $X$, shown on the right,
  with observed data shaded grey.}
  \label{cytofdiagram}
\end{figure}

This setup is illustrated in Figure \ref{cytofdiagram}.  It yields a structure of missingness in the data just as we have
described above.  Each batch of cells corresponds to a group of observations
$G_1,\dots,G_K$, while each set of probes and resulting measurements
corresponds to a set of observable variables $J_1,\dots,J_K$.

As experiments of this form are only beginning to be run, we use existing data
sets to simulate data of this form.  Evan Newell provided us with the data used
in \cite{cytof}.  We focus on a particular data set of mass cytometry measurements on
proteins from a blood sample of one patient.  Lab procedures were used to
obtain a reasonably homogeneous sample of cells.  The data set then consisted
of measurements of 34 proteins on 43,700 cells.  We divide this data set
randomly into five smaller data sets to give some sense of the variation in the methods
we will apply.  To create missingness patterns
of the desired type, we break each subset into 10 groups, and within each group
we artificially censor 10 of the proteins, leaving measurements of only 24 of the 34 proteins.  Because of the skewness of the measurement distributions, we work with the logarithm of the mass cytometry measurements.

To illustrate the problems that can occur with missing pairs, we
select the top 5\% of correlations from the full dataset, and ensure that the
corresponding pairs of variables never appear together in the same group of
variables.  This means that the corresponding partial correlations will be
under-determined.

To give the EM algorithm a favorable starting point, we begin with the true
covariance matrix for the full data set.  We hold fixed those elements
corresponding to variables which have been observed together, and change the
other elements to place the matrix near the center of the cone (similar to making the
corresponding partial correlations zero).  We do this by minimizing
$\log|\Sigma|$ while holding the identifiable elements fixed.

\begin{figure}[ht!]
  \begin{center}
  \includegraphics[width=\linewidth]{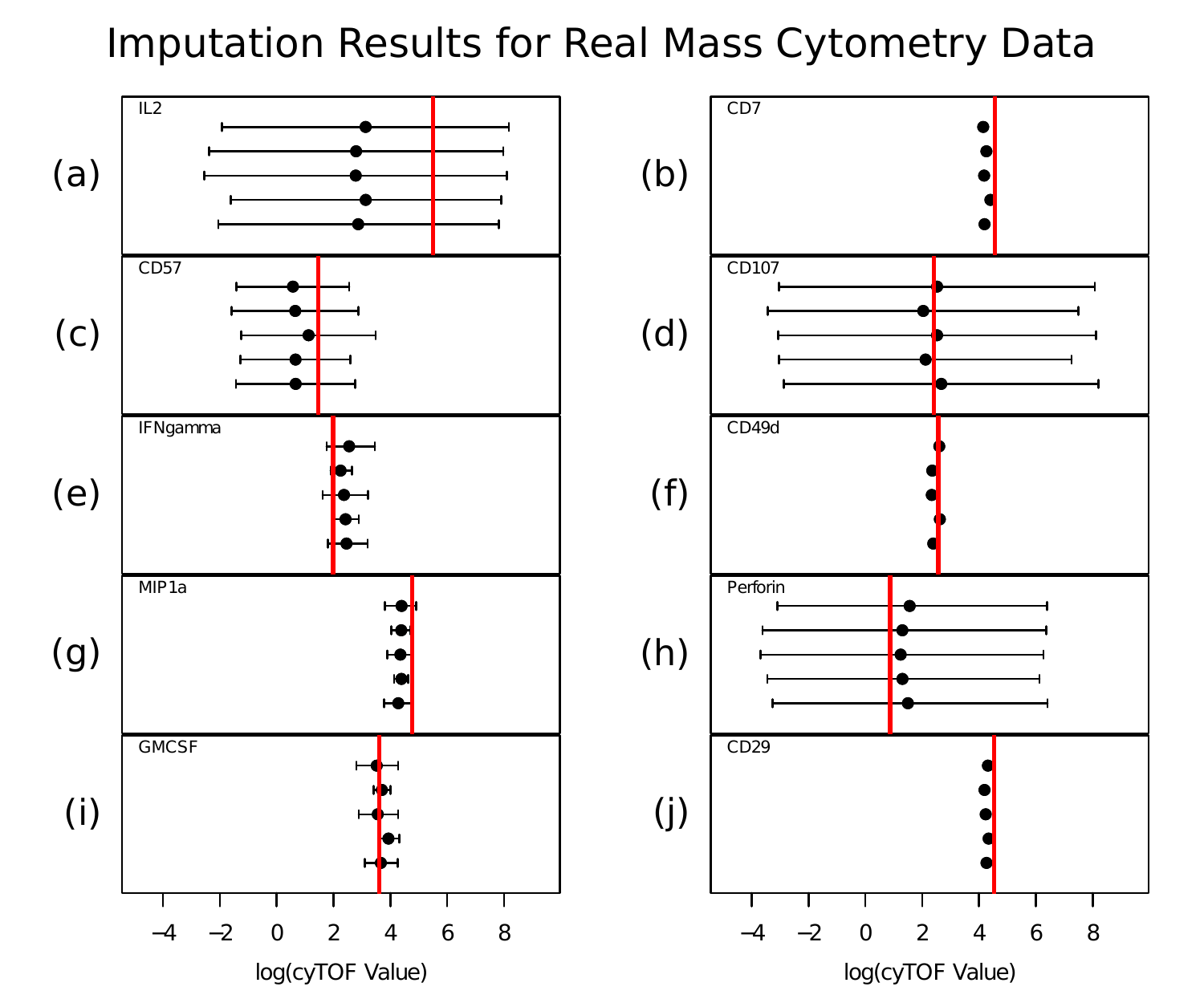}
\end{center}
  \caption{Results for 10 inferred proteins across 5 identically distributed
  synthetic data sets.  Each plot shows the 5 identically distributed realizations.  The imputed EM
  estimates are shown by black dots.  The SDP computed range of possible imputed
  values due to missing partial correlations is shown by the black intervals.  The red vertical line shows the true
  conditional mean, given the full population covariance matrix from the 43,700 completely observed cells.}
  \label{cytofdata}
\end{figure}

For each of these synthetic data sets, we execute both the EM algorithm and our
SDP algorithm.  Figure \ref{cytofdata} shows the resulting estimates for the missing proteins in the first row of $X$.
For comparison,
we show the true conditional means for the missing entries, computed using the
full covariance matrix with no missing data.

In plot $(a)$, we see that the sampling uncertainty does not capture the error
from the true conditional mean.  However, the interval estimates reveal the
potential bias in the estimator.  The error to the true mean falls within this
range of potential bias.

In other cases, like $(b), (f)$ and $(j)$, the intervals have zero width, because
the equivalence set $\Ss$ is degenerate.  This indicates that sampling
uncertainty should be an accurate measure of the error, which it turns out to
be.  Similarly, in $(e),(g),(i)$ and to a lesser extent in $(c)$, we obtain
non-degenerate intervals that have small widths, indicating a small potential
bias.

In cases $(d)$ and $(h)$, our algorithm indicates the potential for very large
biases to be present in the estimator.  Knowing the true mean, we see that this
did not occur.  We expect this conservative behavior in some cases.  Our algorithm
considers the set of possible imputed values over all likelihood-equivalent matrices.
There is no reason to believe that the true unobserved covariance will correspond
to imputed values that are near the edge of all of these intervals.

In this data, we see that it is possible for the lack of identifiability
due to the data structure to lead to errors in imputation that are not captured
by the sampling uncertainty.  In cases where the sampling uncertainty does not
capture the error from the true conditional mean, our interval estimates reveal
the potential for bias in the estimator.  Where our interval estimates are
small, we see that the EM estimates are within sampling uncertainty of the true
value.  This provides evidence that the intervals computed using our algorithm,
though conservative, do capture these errors in imputation due to lack of
identifiability in the data structure.

\section{Discussion}\label{discussion}

This article addresses the issue of missing data imputation by the EM algorithm
in the case where some partial correlations are unobservable, resulting in an
under-identified covariance matrix.  We propose an automatic method for
sensitivity analysis of the resulting imputations to the assumptions being made
on those unobservable correlations.  We present a semidefinite program that
computes the range of possible imputations over the convex set of equal likelihood covariance estimates.
We see from simulation and real biological data that this gives an
accurate sense of the possible errors due to identifying assumptions.

Sensitivity analysis for problems of this nature has been previously discussed
in the literature, including \cite{rubin1986} and \cite{moriarity2001}.
The method in this paper differs in that it
automatically handles the problem of exploring the set of consistent correlation structures, and finds the actual extrema over that set.  To the best of our
knowledge, previous approaches have not automatically explored this space.

The method outlined here could be combined well with methods like
bootstrapping for estimating the uncertainty due to sampling, which would give
intervals that were robust to both sampling uncertainty and identifying
assumptions.  It is also worth noting that our method computes the extrema
separately for each conditional mean to be imputed.  This is bound to be
conservative, as the $\Sigma \in \Ss$ that maximizes or minimizes the
conditional expectation for one coordinate in $X$ is not likely to be correspond to
the extrema of another coordinate.

We saw in the results in Section \ref{results} that the intervals have
the potential to be very wide, if the data were collected in a way that poorly
constrains the important correlations.  This suggests that care should be
taken, when possible,
to design the data missingness to constrain important quantities of
interest.  We believe that the method in this paper could be used to give a better idea of the possible imputation error and to guide the design of the experiments.  Mass cytometry
experiments,
discussed in Section \ref{results:data}, are very amenable to this type of design.  The authors are presently
working on experimental design for mass cytometry data to control the identifiability
issues discussed in this paper.

Finally, much of this paper focuses on bounding the potential error due to partial
identifiability of $\Sigma$ in the case of imputation by conditional
expectation.  It could be interesting to consider other quantities that could
be bounded over the set $\Ss$ in a similar fashion.  The quantities that can be
easily handled are constrained by the need that they be convex over $\Ss$.

This can include functions like Mahalanobis distance and variance
components $u^T\Sigma u$ (for fixed $u$).  This leads to applications
that rely on those measures.  For example, in Principle Component Analysis
(PCA), for a given component $u$, one could bound the range of possible
variances in that direction over the set $\Ss$ by optimizing the affine
function $u^T\Sigma u$.  Similarly, in single-linkage clustering using
Mahalanobis distance, minimizations of $u^T\Sigma^{-1} u$ could be used to give a
lower bound on the separation of the resulting clusters.

\begin{acknowledgements} The authors would like to thank Noah Simon for
  discussion of optimization methods and Jacob Bien for other helpful
  discussions.  We would also like to thank Sean Bendall and Erin Simons for
  first posing the problem to us, and Evan Newell for providing us with the
  mass cytometry data.  
  
  MGG is supported by a National Science Foundation GRFP Fellowship.  SSSO is a
  Taub fellow and is supported by US National Institutes of Health (NIH) (U19
  AI057229).  RT was supported by NSF grant DMS-9971405 and NIH grant N01-HV-28183.  
\end{acknowledgements}

\bibliographystyle{spbasic}

\bibliography{SDPpaperbib}

\appendix
\normalsize
\section{Algorithmic Details}

From Section \ref{proposal:algorithm}, the inner optimization problem we wish
to solve is
\begin{align*}
  \underset{\Sigma}{\mathrm{minimize}} &\;\;
  -\frac{1}{t}\log|\Sigma| + \Sigma_{j,J_k}\hat{\Sigma}_{J_k,J_k}^{-1}(x_{J_k}-\hat{\mu}_{J_k})\\
  \text{subject to } &\;\; \Sigma_{ab} = \hat{\Sigma}_{ab} \text{ for
  $(a,b)\in\bigcup_{\ell} \left(J_{\ell}\times J_{\ell}\right)$}.
\end{align*}

This optimization problem, with fixed $t$, can be solved by generalized gradient descent.  The gradient of the
objective is
\begin{align*}
  \nabla(\mathrm{objective}) &= -\frac{1}{t} \Sigma^{-1} + C\\
  C_{ab} &= \begin{cases}
    \left(\hat{\Sigma}_{J_k,J_k}^{-1}(x_{J_k}-\hat{\mu}_{J_k})\right)[b] & a=j \text{ and } b\in J_k\\
    \left(\hat{\Sigma}_{J_k,J_k}^{-1}(x_{J_k}-\hat{\mu}_{J_k})\right)[a] & a\in J_k \text{ and } b=j\\
    0 & \text{otherwise}
  \end{cases}
\end{align*}

If we initialize the first time with $\Sigma = \hat{\Sigma} \in \Ss$, we want to
remain within $\Ss$ with each step.  Therefore, we project the gradient into
the linear space $\{\Sigma: \Sigma_{ij} = \hat{\Sigma}_{ij} \forall (i,j) \in
\bigcup_k \left(J_k\times J_k\right)\}$.  This is equivalent to holding the
coordinates $(i,j) \in \bigcup_k \left(J_k\times J_k\right)$ fixed, and only taking gradient
steps in the other directions.

With a step size of $\delta$, the update becomes
\begin{align*}
  \Sigma^{(t+1)}_{ab} &= \Sigma^{(t)}_{ab} + \delta\left(\frac{1}{t} (\Sigma^{(t)})_{ab}^{-1} -
  C^{(t)}_{ab}\right) & (a,b) \notin \bigcup_{\ell} \left(J_{\ell}\times J_{\ell}\right)\\%(\text{ where $\hat{\Sigma}_{ab}$ is inestimable})\\
  \Sigma^{(t+1)}_{ab} &= \Sigma^{(t)}_{ab} & (a,b) \in \bigcup_{\ell} \left(J_{\ell}\times J_{\ell}\right)
\end{align*}

Using warm starts, we repeatedly solve this problem with increasing $t$ to
obtain the final solution.  This barrier method is discussed in
\cite{boyd2004convex}, which recommends using a sequence of $t$ that
increase by a factor of $\mu$ (around 10-20) at each outer loop iteration.
More details of the method can be found in Section 11.3.1 of that book.

We
also include acceleration, as in \cite{banerjee2006convex} and
\cite{beck2010gradient}, among others.  This is shown in Figure
\ref{timingplot} to give practically significant improvements in algorithm
timings.  The final algorithm is shown in Algorithm 1.

\begin{algorithm}[t!]
  \DontPrintSemicolon
  \SetKwInOut{Input}{Input}
  \SetKw{Initialize}{Initialize}
  \SetKw{Define}{Define}
  \Input{The estimated covariance matrix $\hat{\Sigma}$}
  \BlankLine
  \Initialize $\Sigma$ and $\Theta$ to $\hat{\Sigma}$\;
  \Initialize $\ell = 1$\;

  \Define $C$ as in Section \ref{proposal:algorithm}\;
  \For{$t = t_0$ \KwTo $t_f$}{
  \Repeat{$\max_{a,b}\left|\Sigma_{ab}-\left(\Sigma_{old}\right)_{ab}\right|<\eps$}{
    $G_{ab} = \frac{1}{t}\left(\Sigma\right)^{-1}_{ab} -
    C_{ab}$ for $(a,b) \notin \bigcup_{\ell} \left(J_{\ell}\times J_{\ell}\right)$\;
      $\Sigma_{old} = \Sigma$\;
      Compute appropriate $\delta$\;
      $\Sigma = \Theta + \frac{\ell}{\ell+3}\left(\Sigma - \delta G - \Theta\right)$\;
      $\Theta = \Sigma_{old} - \delta G$\;
      $\ell = \ell + 1$\;
      \If{$\ell > \ell_{\mathrm{max}}$}{
        $\ell = 1$\;
      }
    }
  }
  \caption{Accelerated algorithm for solving the SDP in Section \ref{proposal:algorithm}.  The variable $\Theta$ is introduced to carry information about the previous steps for the momentum term.  We include restarts in the momentum weight every $\ell_{\mathrm{max}}$ steps, which empirically improves performance.  The procedure for computing an appropriate $\delta$ is shown in Algorithm 2.}
\end{algorithm}

Some care needs to be taken in selecting the step size $\delta$.  We choose it
by backtracking to result in a decrease in the objective, to remain inside the
positive semidefinite cone, and to satisfy the majorization requirements of generalized gradient
descent \citep{beck2010gradient}.  This sub-algorithm is shown as Algorithm 2.

\begin{algorithm}[ht]
  \DontPrintSemicolon
  \SetKwInOut{Input}{Input}
  \SetKwInOut{Output}{Output}
  \SetKw{Initialize}{Initialize}
  \SetKw{Define}{Define}
  \Input{$\Sigma$, $G$}
  \BlankLine
  \Initialize $\delta = 1/0.8$\;
  \Repeat{$\lambda_{\mathrm{min}}\ge 0$ and $f_{obj} \le f_{maj}$}{
    $\delta = 0.8 \delta$\;
    $\Sigma_{\mathrm{test}} = \Sigma - \delta G$\;
    $\lambda_{\mathrm{min}} =
    \min\{\mathrm{eigenvalues}(\Sigma_{\mathrm{test}})\}$\;
    $f_{\mathrm{obj}} = -\frac{1}{t}\log\left|\Sigma_{\mathrm{test}}\right| +
    \left(\Sigma_{\mathrm{test}}\right)_{j,J_k}\Sigma_{J_k,J_k}^{-1}\left(x_{J_k}-\hat{\mu}_{J_k}\right)$\;
    $f_{\mathrm{maj}} = -\frac{1}{t}\log\left|\Sigma\right| +
    \Sigma_{j,J_k}\Sigma_{J_k,J_k}^{-1}\left(x_{J_k}-\hat{\mu}_{J_k}\right) +
    G\circ \left(\Sigma_{\mathrm{test}}-\Sigma\right)$\;
    $\qquad \qquad+\frac{1}{2\delta}\left(\Sigma_{\mathrm{test}}-\Sigma\right)\circ\left(\Sigma_{\mathrm{test}}-\Sigma\right)$\;
  }
  \BlankLine
  \Output{$\delta$}
  \caption{Sub-algorithm to compute appropriate step size $\delta$.  We use $A \circ B$ to denote the Frobenius inner product, $\sum_{a}\sum_b A_{ab}B_{ab}$.}
\end{algorithm}

\end{document}